\documentclass[11pt,a4paper]{article}

\usepackage{authblk}
\usepackage{lineno}
  \usepackage{mathptmx}
\usepackage{subfigure}
\usepackage{sublabel}
\usepackage{dcolumn}
\usepackage{amsmath,amssymb}
\usepackage{bm}
\usepackage{bbm}
\usepackage{color}
\usepackage{overpic}
\usepackage{latexsym}
\usepackage{epstopdf}
\usepackage[english]{babel}
\usepackage{color}
\usepackage{latexsym}
\usepackage{stmaryrd}

\usepackage{dsfont}
\usepackage{psfrag,graphicx} 
\usepackage{epsf} 
\usepackage{subfigure} 
\usepackage{amsmath} 
\usepackage{amssymb} 
\usepackage{amsfonts}
\usepackage{cite}
\usepackage{appendix}

\definecolor{mygrey}{gray}{0.35}
\definecolor{myblue}{rgb}{0.2,0.2,0.8}
\definecolor{myzard}{cmyk}{0,0,0.05,0}
\definecolor{mywhite}{rgb}{1,1,1}
\definecolor{myred}{rgb}{1,0.,0.3}

\usepackage[colorlinks=true,citecolor=myblue,linkcolor=myred]{hyperref}
\def\be{\begin{equation}}
\def\ee{\end{equation}}
\def\ba{\begin{align}}
\def\enda{\end{align}}
\def\bi{\begin{itemize}}
\def\ei{\end{itemize}}

 \def\ee{\mathord{\rm e}}
 
 \def\ii{\mathord{\rm i}}

\def\half{\textstyle\frac{1}{2}}

 \def\ee{\mathord{\rm e}}
 
 \def\ii{\mathord{\rm i}}

\def\half{\textstyle\frac{1}{2}}

\renewcommand{\ii}{{\rm i}}
\renewcommand{\ee}{{\rm e}}

\def\beq{\begin{equation}}
\def\beq{\begin{equation}}
\def\eeq{\end{equation}}

 \newcommand{\ket}[1]{|#1\rangle}
 \newcommand{\bra}[1]{{\langle #1|}}

 \newcommand{\ketbradif}[2]{\ket{#1}\bra{#2}}
 \newcommand{\ketbra}[1]{\ketbradif {#1}{#1}}

 \newcommand{\eq}[1]{Eq.\;(\ref{#1})}

\begin{document}

\title{Noise studies of driven geometric phase gates\\
with trapped ions}
\author[1]{A. Lemmer}
\author[2]{A. Bermudez}
\author[1]{M.~B. Plenio}
\affil[1]{\small Institut f\"{u}r Theoretische Physik, Universit\"{a}t Ulm, 	Albert-Einstein-Allee 11, 89069 Ulm, Germany}
\affil[2]{Instituto de F\'{i}sica Fundamental, IFF-CSIC, Calle Serrano 113b, Madrid E-28006, Spain}

\maketitle

\begin{abstract}
We present a study of the performance of the trapped-ion driven geometric phase gates  introduced in~\cite{driven_bermudez_gates} when realized in a stimulated Raman transition. We show that
the gate can achieve errors below the fault-tolerance threshold in the presence of laser intensity fluctuations. We also find that, in order to reduce the errors due to photon scattering below the fault-tolerance threshold, very intense laser beams are required  to allow for large detunings  in the Raman configuration without compromising the gate speed.
\end{abstract}

\section{Introduction}

Quantum information processing holds the promise of solving efficiently a variety of computational tasks  that are intractable on a classical computer~\cite{NC2010}. Such tasks are routinely decomposed into a series of single-qubit rotations and two-qubit entangling gates~\cite{two_qubit_gates_universal}. While the implementation of  accurate single-qubit gates has been achieved in a variety of platforms~\cite{qc_review}, two-qubit entangling gates with similar accuracies are still very demanding. Such accuracies are compromised by the fact that {\it (i)}   the qubits used to encode the information are not perfectly isolated from the environment, {\it (ii) }  the quantum data bus used to mediate the entangling gates is not perfectly isolated either, and moreover leads to entangling gates that are slower than their one-qubit counterparts,
and {\it (iii)} the tools to process the information  introduce additional external sources of noise. This becomes even more challenging in  light of the   so-called fault-tolerance threshold (FT), which imposes stringent conditions  as these gates  should have errors below  $\epsilon_{\rm FT}=10^{-4}$  for reliable quantum computations~\cite{knill}. Therefore, it is mandatory that two-qubit entangling gates be robust against the typical sources of noise
present in the experiments. This poses an important technological and theoretical challenge. On the one hand,  technology must be improved to minimize all possible sources of noise. On the other hand, theoretical schemes must be devised that minimize the sensitivity of the entangling two-qubit  gates with respect to the most relevant sources of noise.

With trapped ions~\cite{qc_ions}, it is possible to encode a qubit in various manners: there are the so-called ``optical'', ``Zeeman'' and ``hyperfine'' qubits. Here, we shall focus on hyperfine qubits. In this approach, the
qubit states are encoded in two hyperfine levels of the electronic ground-state manifold, and the qubit transition frequency typically lies in the microwave domain.  Hyperfine qubits offer the advantage that spontaneous
emission from the qubit levels is negligible, in practice. Additionally, one-qubit gates can be implemented via microwave radiation, which has already been shown to allow for errors below the FT~\cite{1_qubit_mw_gates}.

Entangling two-qubit gates require   a quantum data bus to mediate the interaction between two distant qubits. The most successful schemes in trapped ions~\cite{zoller_cirac, ms_gate1, didi_gate} make use of the collective 
vibrations of the ions in a harmonic trap to mediate interactions between the qubits. The more recent {\it driven geometric phase gate}~\cite{driven_bermudez_gates,far_detuned_driven_geometric_phase_gate,nist_ss_gate}, which is the subject of this work,  also relies on phonon-mediated interactions and  thus requires a qubit-phonon coupling.
In the case of hyperfine qubits, the qubit-phonon coupling is not easily provided with microwave radiation.
Although there are schemes to achieve such a coupling  by means of magnetic field gradients~\cite{static_gradients, oscillating_gradients}, spin-phonon coupling is most commonly  provided
by optical radiation in a so-called stimulated Raman configuration. In this setup, transitions between the qubit levels are off-resonantly driven via a third auxiliary level from the excited state manifold by a pair of laser beams. Therefore, in contrast to  the direct
microwave coupling, spontaneous photon emission may occur, which acts as an additional source of noise with detrimental effects on the 
gate performance~\cite{plenio_lower_bounds, plenio_lower_bounds_proc}.

In this manuscript, we will complement the analysis of  the driven geometric phase gate in the presence
of noise~\cite{driven_bermudez_gates}, where we showed its built-in resilience to thermal fluctuations, dephasing noise, and drifts of the laser phases. There, we also explored the behavior of the gate with respect to  microwave intensity noise, and proposed ways to increase its robustness. In this manuscript, we consider two additional sources of noise that are present in experiments, namely laser intensity fluctuations and residual spontaneous emission.
The first part of the manuscript is devoted to the study of the stimulated Raman configuration, and the derivation of an effective dynamics within the qubit manifold using the formalism of~\cite{sorensen_reiter}. 
This allows us to obtain expressions for the desired qubit-phonon coupling and the residual spontaneous emission. We then use these expressions  to analyze the effects of photon scattering by numerically simulating the gate dynamics in such a stimulated Raman
configuration. Subsequently, we investigate the performance of the gate in the presence of laser intensity fluctuations. Finally, in the last section we provide a summary of the results of this manuscript.

\section{The setup}

\subsection{The stimulated Raman configuration}

\begin{figure}[hbt]
\centering
\includegraphics[width=0.8\textwidth]{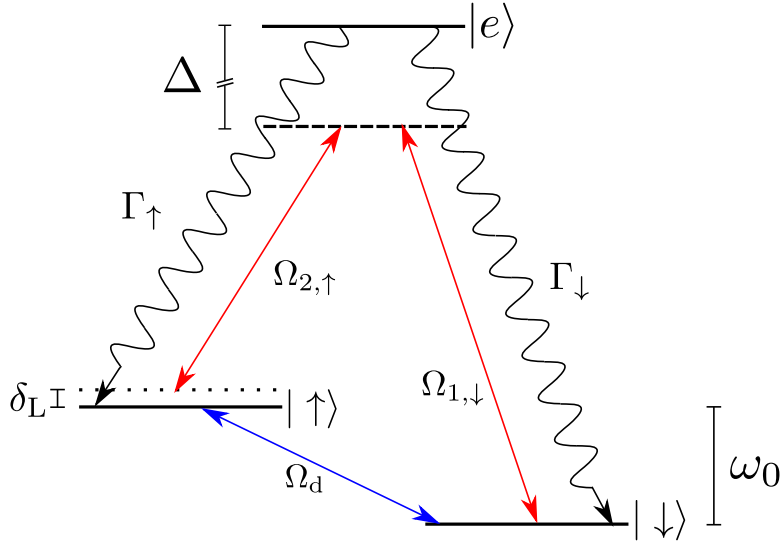}
\caption{{\bf Stimulated Raman transition:} Two far-detuned lasers (red arrows) are applied to a $\Lambda-$type system. Laser $L_1$ couples to the transition $\ket{\downarrow}\leftrightarrow\ket{e}$ with Rabi frequency
$\Omega_{1,\downarrow}$ and $L_2$ couples to $\ket{\uparrow}\leftrightarrow\ket{e}$ with Rabi frequency $\Omega_{2,\uparrow}$. The beatnote of the two lasers is tuned close to the qubit transition frequency 
$\omega_1-\omega_2 = \omega_0 +\delta_{\rm L},\:\delta_{\rm L}\ll \omega_0$. Both beams are detuned by an amount $\Delta$ from the respective transition. By choosing $\Delta$ much larger than the excited state's
linewidth $\Gamma=\Gamma_{\uparrow}+\Gamma_{\downarrow}$ and the individual Rabi frequencies effective two-level dynamics within the qubit manifold arise. \newline
{\bf Driven geometric phase gate:} In order to achieve the driven geometric phase gate two lasers in a stimulated Raman configuration are tuned close to the first red-sideband excitation of the phonons in $x-$direction
$\omega_1-\omega_2 = (\omega_0-\omega_n) + \delta_n,\:\delta_n \ll \omega_n$. Additionally, a strong direct microwave drive with Rabi frequency $\Omega_{\rm d}$ is applied on the carrier transition (blue arrow). Spontaneous
emission from the auxiliary level is indicated by the curly lines.}
\label{fig1}
\end{figure}

Let us consider the situation depicted in Fig.~\ref{fig1}. For the moment, we will neglect the fact that we are dealing with ions in a harmonic trap. We consider a $\Lambda-$type 
three-level system that is illuminated by two lasers $L_1$ and $L_2$ with frequencies $\omega_1$ and $\omega_2$, respectively. The levels $\ket{\downarrow}$ and $\ket{\uparrow}$  
form the qubit. We denote the qubit transition frequency by $\omega_0=\epsilon_{\uparrow} - \epsilon_{\downarrow}$ where $\epsilon_s$ is the energy of state $\ket{s}$ .
Note that we  set $\hbar = 1$  throughout this manuscript. The beatnote of the two lasers is tuned close to the qubit transition frequency $\omega_1 - \omega_2 \approx \omega_0$.
We assume that each of the laser beams only couples one of the qubit levels to the excited state $\ket{e}$, and is detuned by an amount $\Delta$ from the respective transition. Here, we consider that  $L_1$ only 
couples to the transition $\ket{\downarrow}\leftrightarrow \ket{e}$ with Rabi frequency $\Omega_{1,\downarrow}$ and $L_2$ only to $\ket{\uparrow} \leftrightarrow \ket{e}$ with Rabi frequency $\Omega_{2,\uparrow}$. Hence, the Hamiltonian of the system is given by
\beq
H_{\rm full} =\hspace{-1ex}  \sum\limits_{s=\downarrow,\uparrow,e}\hspace{-1ex} \epsilon_s \ketbra{s} + \half\left( \Omega_{1,\downarrow}\ee^{\ii({\bf k}_1 \cdot {\bf r} - \omega_1 t + \varphi_1)} \ketbradif{e}{\downarrow}
+ \Omega_{2,\downarrow}\ee^{\ii({\bf k}_2 \cdot {\bf r} - \omega_2 t + \varphi_2)} \ketbradif{e}{\uparrow} +{\rm H.c.}\right)
\eeq
where ${\bf k}_{1/2}$ and $\varphi_{1/2}$ are the laser wave vectors and phases, and ${\bf r}$ is the position of the ion.

Choosing the detuning much larger than the auxiliary state's linewidth $\Delta \gg \Gamma$ and the individual Rabi frequencies $\Delta \gg |\Omega_{1,\downarrow}|,\:|\Omega_{2,\uparrow}|$, the excited state population is
very small and can be adiabatically eliminated from the dynamics. Applying the formalism presented in~\cite{sorensen_reiter}, we obtain the effective Hamiltonian
\begin{equation}
\begin{split}
H_{\rm eff} &= (\epsilon_{\downarrow} +\Delta\epsilon_{\downarrow} ) \ketbra{\downarrow} +
(\epsilon_{\uparrow} +\Delta\epsilon_{\uparrow} ) \ketbra{\uparrow} -
\left(\frac{\Omega_{1,\downarrow} \Omega_{2,\uparrow}^*}{4 \Delta} \sigma^+ \ee^{\ii({\bf k}_{\rm L}\cdot{\bf r}- \omega_{\rm L} t)} + {\rm H.c.} \right) 
\end{split}
\label{raman_eff_ham_complete}
\end{equation}
within the qubit manifold, where $\Delta\epsilon_{\downarrow}=-|\Omega_{1,\downarrow}|^2 /4\Delta$, and $\Delta\epsilon_{\uparrow}= - |\Omega_{2,\uparrow}|^2 /4\Delta$. Here ${\bf k}_{\rm L}={\bf k}_{1}-{\bf k}_{2}$ (${\omega}_{\rm L}={\omega}_{1}-{\omega}_{2}$) denotes the effective laser wave vector (frequency) and we have set the 
Raman-beam phase $\varphi_{\rm L}=\varphi_1 -\varphi_2 = 0$. Furthermore, we have introduced $\sigma^+ = \ketbradif{\uparrow}{\downarrow}=(\sigma^-)^{\dagger}$ and
define $\sigma^z = \ketbra{\uparrow} - \ketbra{\downarrow}$.
After moving to an interaction picture with
respect to the ground-state Hamiltonian $H_{\rm g} = \half\omega_0\sigma^z$ the Hamiltonian \eq{raman_eff_ham_complete} becomes
\beq
H_{\rm int} = \Delta\epsilon_{\downarrow} \ketbra{\downarrow} +\Delta\epsilon_{\uparrow} \ketbra{\uparrow}
+\left(\frac{\Omega_{\rm L}}{2} \sigma^+ \ee^{\ii({\bf k}_{\rm L}\cdot{\bf r}- \delta_{\rm L} t)} + {\rm H.c.} \right)
\label{raman_eff_ham_ip}
\eeq
where $\delta_{\rm L}=(\omega_1 -\omega_2) -\omega_0$  denotes the detuning of the two lasers' beatnote from the qubit transition frequency, such that $|\delta_{\rm L}| \ll \omega_0$, and we have introduced the effective
Rabi frequency of the Raman transition
\beq
\Omega_{\rm L}=-\frac{\Omega_{1,\downarrow}\Omega_{2,\uparrow}^*}{2\Delta}.
\label{eff_rabi_freq}
\eeq
The first part of \eq{raman_eff_ham_ip} represents the ac-Stark shifts of the qubit levels in the presence of the two laser fields. In general, the ac-Stark shifts are different for the two levels, such that the differential ac-Stark shift leads to dephasing when the laser intensities fluctuate.
Therefore, they are typically  nulled in experiments by choosing the right laser polarizations and intensities. In this manuscript,  we  assume $|\Omega_{1,\downarrow}|=|\Omega_{2,\uparrow}|$ such that 
the ac-Stark shifts can be neglected~\cite{comment_ac_ss}. From the second part of~\eq{raman_eff_ham_ip} we can see that we obtain a coupling between the qubit levels with an effective Rabi frequency given by~\eq{eff_rabi_freq}.
Note that the effective wave vector in~\eq{raman_eff_ham_ip} is an optical wave vector and will lead to a non-negligible Lamb-Dicke factor and thus enable a qubit-phonon coupling.

\subsection{Spontaneous emission}

So far, we have only provided the coherent Hamiltonian dynamics and neglected incoherent scattering processes. However, elastic and inelastic photon scattering (i.e. spontaneous emission) may occur in this configuration
as indicated by the curly lines in Fig.~\ref{fig1}. In the Schr\"{o}dinger picture, the complete dynamics is described by the master equation~\cite{master_eq_review_susana}
\beq
\dot{\rho} = -{\rm i}[H_{\rm full},\rho] +  \sum_{s=\uparrow,\downarrow}\left( L_{s} \rho L_{s}^{\dagger} -\half\{L_{s}^{\dagger} L_{s} , \rho  \} \right)
\label{master_eq}
\eeq
where
the incoherent photon-scattering processes are described by the jump operators
\beq
L_{\downarrow} = \sqrt{\Gamma_{\downarrow}} \ketbradif{\downarrow}{e},\hspace{3ex} L_{\uparrow} = \sqrt{\Gamma_{\uparrow}} \ketbradif{\uparrow}{e},
\label{lindblads_full}
\eeq
where $\Gamma_{\downarrow}$ and $\Gamma_{\uparrow}$ denote the scattering rates to the levels $\ket{\downarrow}$ and $\ket{\uparrow}$, respectively. The total scattering rate is then given by 
$\Gamma = \Gamma_{\downarrow} + \Gamma_{\uparrow}$. Starting from one of the qubit levels two processes may occur: a photon is scattered elastically, such that we end up in the same qubit state but the
state has acquired a random phase. This process is known as Rayleigh scattering. Besides, it may happen that a photon is scattered inelastically and the qubit state is changed. This process is termed
Raman scattering. It is clear that both of the processes will have a detrimental effect on the entangling gate.
Since we seek an effective dynamics within the qubit manifold, we again use the formalism of~\cite{sorensen_reiter}, to obtain the effective master equation
\beq
\dot{\rho} = -{\rm i}[H_{\rm eff},\rho] +  \sum_{s=\uparrow,\downarrow}\bigg( L_{s}^{\rm eff} \rho (L_{s}^{\rm eff})^{\dagger} -\half\{(L_{s}^{\rm eff})^{\dagger} L_{s}^{\rm eff} , \rho  \} \bigg),
\label{eff_master_eq}
\eeq
where the effective Hamiltonian is given in \eq{raman_eff_ham_ip}, and the effective jump operators correspond to 
\begin{eqnarray}
L_{\downarrow}^{\rm eff} &=& \frac{\Omega_{1,\downarrow} \sqrt{\Gamma_{\downarrow}}}{2\Delta - \ii \Gamma} \ee^{\ii ({\bf k}_1 \cdot {\bf r} - \omega_{1,\downarrow}  t)} \ket{\downarrow}\bra{\downarrow}
+ \frac{\Omega_{2,\uparrow} \sqrt{\Gamma_{\downarrow}}}{2\Delta - \ii \Gamma} \ee^{\ii({\bf k}_2 \cdot {\bf r}-  \omega_{2,\uparrow}  t)} \ket{\downarrow}\bra{\uparrow}, \label{jump_down} \\
L_{\uparrow}^{\rm eff} &=& \frac{\Omega_{1,\downarrow} \sqrt{\Gamma_{\uparrow}}}{2\Delta - \ii \Gamma} \ee^{\ii({\bf k}_1 \cdot {\bf r} - \omega_{1,\downarrow}  t)} \ket{\uparrow}\bra{\downarrow}
+ \frac{\Omega_{2,\uparrow} \sqrt{\Gamma_{\uparrow}}}{2\Delta - \ii \Gamma} \ee^{\ii({\bf k}_2 \cdot {\bf r} - \omega_{2,\uparrow}  t)} \ket{\uparrow}\bra{\uparrow}, \label{jump_up}
\end{eqnarray}
where we have introduced the frequencies $\omega_{l,s}=\omega_l + \epsilon_s$. From the above equations it follows that the effective decay rates scale as 
$\Gamma_s^{\rm eff} \propto \Gamma_s |\Omega_{l,s}|^2/ \Delta^2$. Thus, it is clear that the effective decay rates can be suppressed by increasing the detuning $\Delta$. From
\eq{eff_rabi_freq} we see that increasing $\Delta$ demands an increase in the modulus of the individual Rabi frequencies and thus in the applied laser power to maintain the value of
the effective Rabi frequency, which will in turn determine the gate speed.

\subsection{Spin-motion coupling}

In this part, we want to extend our analysis of the stimulated Raman configuration and derive the spin-motion coupling. To this end, we now consider a string of $N$ ions in a linear Paul trap. 
We assume that the ions are sufficiently cold such that their motion, which is strongly coupled via the Coulomb interaction, is described in terms of normal modes~\cite{james}. For a crystal of $N$ ions there
are $N$ normal modes in every spatial direction, and thus the motional Hamiltonian is given by
\beq
H_{\rm m} =  \sum_{\alpha=x,y,z} \sum_{n=1}^{N} \omega_{\alpha,n} a^{\dagger}_{\alpha,n} a_{\alpha,n}
\label{motional_ham}
\eeq
where $\omega_{\alpha,n}$ denotes the motional mode frequency of mode $n$ in spatial direction $\alpha$ and $ a^{\dagger}_{\alpha,n} \: (a_{\alpha,n})$ is the respective creation (annihilation) operator. 
As we can see from~\eq{motional_ham}, the normal modes of different spatial directions are uncoupled. Assuming the ions are illuminated in a stimulated Raman configuration the system's full Hamiltonian reads
\beq
H_{\rm full} = H_{\rm m}+ \sum\limits_{i,s}  \epsilon_s \ket{s}_i \bra{s} 
+ \frac{1}{2}\sum\limits_{i}\left( \Omega_{1,\downarrow}\ee^{\ii({\bf k}_1 \cdot {\bf r}_i - \omega_1 t)} \ket{e}_i \bra{{\downarrow}} + \Omega_{2,\uparrow}\ee^{\ii({\bf k}_2 \cdot {\bf r}_i - \omega_2 t)} \ket{e}_i \bra{{\uparrow}} +{\rm H.c.}\right)
\label{full_ham_lambda}
\eeq
where $i=1,\ldots,N$ denotes the individual ions. Again, by performing adiabatic elimination of the excited state, we obtain the Hamiltonian
\beq
H_{\rm eff}= H_0 + \sum_i \left(\frac{\Omega_{\rm L}}{2} \sigma_i^+ \ee^{\ii({\bf k}_{\rm L}\cdot{\bf r}_i- \omega_{\rm L} t)} + {\rm H.c.} \right)
\label{stimulated_raman_phonons}
\eeq
where we have defined
\beq
H_0 = \sum_i \frac{\omega_0}{2} \sigma^z_i + \sum_{\alpha,n} \omega_{\alpha,n} a^{\dagger}_{\alpha,n} a_{\alpha,n}.
\label{def_h0}
\eeq
We will now assume that the effective laser wave vector is 
aligned along the $x$-axis of the trap, i.e. ${\bf k}_{\rm L} = k_{\rm L} {\bf e}_x$. In this case, the laser only couples to the phonons in $x-$direction and we will omit the index $\alpha$ as it is
clear that we refer to the case $\alpha=x$. We can now rewrite the $x$-component of the ion's position as
\beq
x_i =x_i^0 + \sum_{n=1}^N M_{in}\frac{1}{\sqrt{2m\omega_n}} (a_n +a_n^{\dagger})
\eeq 
where $M_{in}$ is the dimensionless normalized amplitude of mode $n$ for ion $i$ and $x_i^0$ is the equilibrium position of ion $i$ in $x$-direction. Assuming the trap axis
along the $z-$direction, we have $x_i^0=0$ for all ions. Using the Lamb-Dicke parameter (LDP) $\eta_n=k_{\rm L}/\sqrt{2m\omega_n}$ and moving to an
interaction picture with respect to $H_0$,~\eq{stimulated_raman_phonons} can be 
recast into the form
\beq
H_{\rm eff}= \sum_i \frac{\Omega_{\rm L}}{2} \ee^{\ii\sum_n M_{in} \eta_n (a_n\ee^{- \ii \omega_n t} +a_n^{\dagger} \ee^{ \ii \omega_n t})} \sigma_i^+ 
\ee^{-\ii \delta_{\rm L} t} + {\rm H.c.}.
\label{stimulated_raman_eta}
\eeq
Typically, for optical wave vectors we have $\eta_n \simeq 0.1$ such that we can expand the first exponential in the above equation in powers of $\eta_n$. Assuming a laser detuning $\delta_{\rm L}=-\omega_n +\delta_n,\:\delta_n\ll\omega_n$, we obtain the red-sideband excitation, and~\eq{stimulated_raman_eta} finally becomes
\beq
H_{\rm rsb} = \sum_{i,n} \mathcal{F}_{in} \sigma_i^+ a_n \ee^{-\ii \delta_n t} + {\rm H.c.}.
\label{h_eff_final_rsb}
\eeq
Here we performed a rotating wave approximation relying on $|\Omega_{\rm L}| \ll \omega_n$ and introduced the sideband coupling strengths
\beq
\mathcal{F}_{in}=\ii \frac{1}{2}\eta_n \Omega_{\rm L} M_{in}.
\label{def_sideband_coupling_strengths}
\eeq
In the following we shall consider $N=2$ ions, in this case $M_{12}=-1/\sqrt{2}$ and $M_{in}=1/\sqrt{2}$ else.

\section{Driven geometric phase gates with trapped ions}\label{theory_chapter}

In this section we will briefly explain the driven geometric phase gate introduced in~\cite{driven_bermudez_gates}. We start by considering a crystal of $N=2$ trapped ions in a linear Paul trap.
The internal degrees of freedom of the ions are treated in the usual two-level approximation, and we again assume that the motion of the ions is described in terms of normal modes. 
Thus, the system is described by the Hamiltonian $H_0$ in ~\eq{def_h0}. Additionally, a strong microwave that drives the qubit transition directly is applied to the ions.
The interaction of the ions with the microwave field is described by the Hamiltonian
\beq
H_{\rm d} = \sum_{i=1}^N \frac{\Omega_{\rm d}}{2} \sigma_i^+ \ee^{- \ii \omega_{\rm d} t} + {\rm H.c.}
\label{def_hd}
\eeq
where $\omega_{\rm d}$ ($\Omega_{\rm d}$) denotes the (Rabi) frequency of the applied microwave, and we have assumed that $|\Omega_{\rm d}|\ll\omega_{\rm d}\sim\omega_0$.

In order to realize an interaction between the internal states of the ions, we use the coupled harmonic motion of the ions as was done in previous schemes.
The coupling between the internal states of the ions and their motion is provided by two-photon stimulated Raman transitions via a third auxiliary level (see above).
The Hamiltonian describing the qubit phonon coupling is then given by
\beq
H_{\rm qp} = \sum_i \frac{\Omega_{\rm L}}{2} \sigma_i^+ \ee^{\ii({\bf k}_{\rm L}\cdot{\bf r}_i- \omega_{\rm L} t)} + {\rm H.c.}
\eeq
where all quantities were already introduced. The system's full Hamiltonian is then given by
\beq
H_{\rm dss} = H_{0} + H_{\rm d} + H_{\rm qp}.
\label{full_gate_ham}
\eeq
Recall that we have assumed equal Rabi frequencies for the two lasers such that ac-Stark shifts are negligible. Let us briefly remark at this point that a full analysis starting from
the complete three level system illuminated by the Raman lasers and the microwave gives the same effective Hamiltonian in the limit $\Delta \gg \Omega_{\rm d},|\Omega_{l,s}|,\delta_{\rm L}$. 
This limit will be fulfilled in all our considerations here. 

Setting the laser frequencies such that $\omega_{\rm L}= (\omega_0 - \omega_n)+\delta_n$, where $\delta_n \ll \omega_n$ and $|\Omega_{\rm L}|\ll\omega_n$, we obtain the first red-sideband
excitation as was explained above. 
Moving to an interaction picture with respect to $H_0$ and assuming that the microwave is on resonance with the qubit transition $\omega_{\rm d}=\omega_0$, the interaction of the ions with
the applied fields is described by the ``driven single sideband'' Hamiltonian
\begin{equation}
\tilde{H}_{\rm dss}=\sum_i \frac{\Omega_{\rm d}}{2} \sigma^x_i + \sum \limits_{i,n} \left[\mathcal{F}_{in} \sigma_i^+ a_n \ee^{-\ii \delta_n t} + {\rm H.c.}\right].
\label{qubit_phonon}
\end{equation}
The situation is depicted in Fig.~\ref{fig1}. Note that we performed two rotating wave approximations here using $|\Omega_{\rm d}|/\omega_0 \ll 1$ and $|\Omega_{\rm L}|/\omega_n\ll 1$. 
Moving to yet another interaction picture with respect to the applied microwave driving term, the qubit-phonon interaction Hamiltonian reads
\begin{equation}
\begin{split}
\tilde{\tilde{H}}_{\rm dss}= \sum \limits_{i,n}\! \frac{\mathcal{F}_{in}}{2} (\!\sigma_i^x+\ii\sigma_i^y\cos(\Omega_{\rm d}t)-\ii\sigma_i^z\sin(\Omega_{\rm d}t)\!) a_n \ee^{-\ii \delta_n t} + {\rm H.c.}.
\end{split}
\label{single_sideband_ip2}
\end{equation}
In the above equation we can observe that three state-dependent forces in the $\sigma^x,\:\sigma^y$ and $\sigma^z$ bases act on the ions. However, the forces in $\sigma^y$ and $\sigma^z$ rotate at
the Rabi frequency of the microwave driving while the force in $\sigma^x$ is left unaltered. Now, if the microwave driving is sufficiently strong $|\Omega_{\rm d}| \gg |\mathcal{F}_{in}|, \delta_n$, we can neglect the $\sigma^y$ and $\sigma^z$ 
forces in a rotating wave approximation, such that we are left with a single state-dependent force in the $\sigma^x$-basis. 
In this limit, we can approximate the Hamiltonian in~\eq{single_sideband_ip2} as
\begin{equation}
\label{x_force}
\tilde{\tilde{H}}_{\rm dss}\approx\sum \limits_{i,n}\half \mathcal{F}_{in} \sigma_i^x a_n \ee^{-\ii \delta_n t} + {\rm H.c.}.
\end{equation}
We should note that for trapped ions there have been different proposals using strong-driving assisted gates~\cite{light_shift_gates, light_shift_gates_thermal}.
This has also been considered in the context of cavity-QED to generate entangled states of different cavity modes~\cite{strong_driving_cavities_solano}, or atomic entangled states in thermal cavities~\cite{strong_driving_cavities_zheng}.

\begin{figure}
\centering
\includegraphics[width=0.4\textwidth]{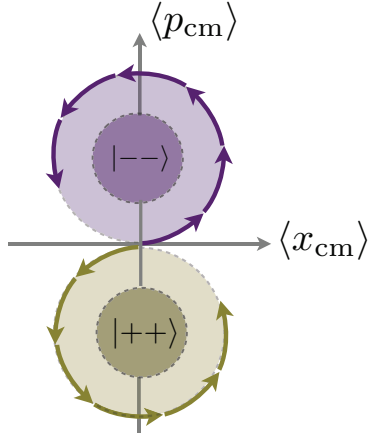}
\caption{{\bf Phase space trajectories.} The figure illustrates the action of the Hamiltonian in~\eq{x_force} on the center-of-mass mode if the internal states of the ions
are in the states $\ket{++}$ or $\ket{--}$, respectively. The mode is displaced along circular trajectories in phase space as indicated by the arrows. If the trajectories
are closed, the motion returns to its initial state and the full state consisting of the ions internal and motional states acquires a phase equal to the area enclosed in 
phase space.}
\label{fig_phase_space_trajectories}
\end{figure}

Let us briefly analyze the action of the Hamiltonian in~\eq{x_force}, which consists of a single state-dependent force in the $\sigma^x$-basis that aims at displacing the
normal modes of the ions in $x$-$p$ phase space. If the ions are in an eigenstate of $\sigma_1^x \sigma_2^x$, the above Hamiltonian couples to one of the normal modes and
displaces it along periodic circular trajectories in phase space. This situation is illustrated in Fig.~\ref{fig_phase_space_trajectories}. 
In general, this Hamiltonian will produce entanglement between the internal and motional states of the ions.
Yet, if the internal and motional states of the ions are in a product state initially, and the phase-space trajectories are closed, the internal and motional states end up in a 
product state again. In this case the state of the system acquires a phase which is equal to the area enclosed in phase space~\cite{didi_gate}. 
In our setup, the eigenstates $\ket{++}_x$ and $\ket{--}_x$ couple to the center-of-mass mode, while the eigenstates $\ket{+-}_x$ and $\ket{-+}_x$ couple to the zig-zag mode. Since the areas enclosed
in phase space are different for the two modes, one can adjust the parameters such that, up to an irrelevant global phase, we obtain the following truth table
\begin{equation}
\label{table_phase_gate}
\begin{split}
&\textstyle{\ket{++}_x\to\phantom{\ee^{\ii\frac{\pi}{2}}}\ket{++}_x,}\\
&\textstyle{\ket{+-}_x\to\ee^{\ii\frac{\pi}{2}}\ket{+-}_x,}\\
&\textstyle{\ket{-+}_x\to\ee^{\ii\frac{\pi}{2}}\ket{-+}_x,}\\
&\textstyle{\ket{--}_x\to\phantom{\ee^{\ii\frac{\pi}{2}}}\ket{--}_x.}\\
\end{split}
\end{equation} 
This is a conditional $\pi/2$ phase gate, which is an entangling gate in the usual computational basis. Together with single-qubit rotations, it forms a universal set of gates for quantum computation. 
The major asset of the geometric phase gates is that they are {\it insensitive} to the ions motional state in case the trajectories are closed and internal and motional states were in a product state initially.
Note that for weak drivings $\Omega_{\rm d}$, the approximation leading to~\eq{x_force} is not valid any more, the phase-space trajectories do not close, and the
geometric character of the gate is lost.

Remarkably, the unitary evolution of the  Hamiltonian in~\eq{x_force} is exactly solvable. If the trajectories are closed, i.e. $t_{\rm g}=k_n 2\pi/\delta_n,\:k_n \in \mathbb{Z}$, the time-evolution operator
generated by the Hamiltonian in~\eq{x_force} is given by
\begin{equation}
\hat{U}(t_{\rm g})=\ee^{-\ii t_{\rm g}\sum_{ij}J_{ij}^{\rm dss}\sigma_i^x\sigma_j^x},\hspace{2ex} J_{ij}^{\rm dss}=\textstyle{-\sum_n\frac{\mathcal{F}_{in}\mathcal{F}^*_{jn}}{4\delta_n}}.
\label{ms_gate}
\end{equation}
It is readily checked that for $t_g (2J_{12}^{\rm dss})=\pi/4$, this yields the truth table in~\eq{table_phase_gate}. 

Unfortunately, due to the different non-commuting qubit-phonon couplings in~\eqref{single_sideband_ip2}, the associated unitary evolution operator cannot be calculated exactly. However, the leading-order contributions for strong drivings can be obtained by means of a Magnus expansion~\cite{driven_bermudez_gates}. This allowed us to derive  a set of conditions under which the approximation in~\eq{x_force} for the Hamiltonian~\eqref{single_sideband_ip2} is fulfilled. We showed, both analytically and numerically, that microwave drivings with  strengths $\Omega_{\rm d}$ of a few MHz typically suffice. It was also shown
that the constraint $t_{\rm g}=k_n 2\pi/\delta_n,\:k_n \in \mathbb{Z}$ can be fulfilled such that the geometric character is fulfilled for general two-qubit states. Moreover, when complemented with a single spin-echo pulse, we showed that the gate is extremely well approximated by~\eq{ms_gate}.

\section{Gate performance in the presence of noise}\label{noise_chapter}

Once  the mechanism of the driven geometric phase gates has been described, we are interested in its performance in the presence of typical sources of noise in ion traps. We showed in~\cite{driven_bermudez_gates}  that the gate can attain
errors below the FT in the presence of residual thermal motion, laser phase and dephasing noise. Also, a constraint on the intensity stability of the microwave driving was given. However, the impact of spontaneous emission
and laser intensity noise was not investigated. In this section, we shall study the effects of these two additional sources of noise. 

\subsection{Spontaneous emission}

In order to include the effects of spontaneous emission, we study the gate dynamics in a master equation approach. For our numerical simulations, we used a realistic set of parameters for an ion-trap experiment summarized in
Table~\ref{tab_1}, assuming the Raman beams to be at a right angle. As a specific ion species we chose $^{25}$Mg$^{+}$. In order to perform the simulations as efficiently as possible we cast the Hamiltonian $H_{\rm dss}$ in~\eq{full_gate_ham}
in a picture where it becomes time independent, namely
\begin{equation}
\begin{split}
H_{\rm dss}'=& U(t)H_{\rm dss}U(t)^{\dagger} \\
=&\sum_n\delta_na_n^{\dagger}a_n+\sum_i\frac{\Omega_{\rm d}}{2}\sigma_i^x+\sum_{i,n}(\mathcal{F}_{in}\sigma_i^+a_n+\text{H.c.}),
\end{split}
\label{dss_time_indep}
\end{equation}
where the unitary transformation is given by ${U}(t)=\ee^{\ii t\sum_n(\omega_0-\omega_{\rm L})a_n^{\dagger}a_{n}}\ee^{\ii t \sum_i\frac{1}{2}\omega_0\sigma_i^z}$. The effective Lindblad operators in this picture 
are given by Eqs.~\eqref{jump_down} and~\eqref{jump_up}. These operators are time-dependent and only appear in the form $(L_k^{\rm eff})^{\dagger} L_k^{\rm eff}$ and $L_k^{\rm eff} \rho (L_k^{\rm eff})^{\dagger}$.
Analyzing these products  in detail, we find that there are time-independent terms, but also additional  terms carrying oscillatory time dependences with frequencies $\sim \delta_{\rm L},\omega_0$. Since the amplitude of these terms fulfills $\Gamma_s |\Omega_{l,s}|^2/\Delta^2 \ll \delta_{\rm L},\omega_0$ for the constraints taken so far,
we can safely neglect their contributions in the time evolution. The effective dissipative processes are then described by the jump operators
\beq
\begin{split}
& L^{\rm eff}_{\downarrow \downarrow}=\sqrt{\frac{\Gamma_{\downarrow}}{4 \Delta^2 + \Gamma^2} \frac{|\Omega_{1,\downarrow}|^2}{2}} \sigma^z, 
\qquad L^{\rm eff}_{\uparrow \downarrow} = \sqrt{\frac{\Gamma_{\downarrow}}{4 \Delta^2 + \Gamma^2} |\Omega_{2,\uparrow}|^2} \sigma^-, \\
& L^{\rm eff}_{\downarrow \uparrow}=\sqrt{\frac{\Gamma_{\uparrow}}{4 \Delta^2 + \Gamma^2} |\Omega_{1,\downarrow}|^2} \sigma^+,
\qquad L^{\rm eff}_{\uparrow \uparrow} = \sqrt{\frac{\Gamma_{\uparrow}}{4 \Delta^2 + \Gamma^2} \frac{|\Omega_{2,\uparrow}|^2}{2}} \sigma^z. \\
\end{split}
\eeq
Note, the Lindblad operators $\propto \sigma^z$ describe dephasing caused by Rayleigh scattering where the qubit state is not altered. The remaining contributions describe Raman
scattering where the qubit state is altered upon a scattering event. For our simulations we assume equal branching ratio for the two decay channels, i.e. $\Gamma_{\downarrow}=\Gamma_{\uparrow}=\Gamma/2$.

\begin{table}[b]
\centering
  \caption{{\bf Values of trapped-ion setup for the numerical simulation and parameters of $^{25}$Mg$^+$}  }
\begin{tabular}{ c  c c  c  c  c  c  c c}
\hline \hline
  $\omega_z/2 \pi$ & $\omega_x/2 \pi$ & $\delta_{{\rm com}}/2\pi$ &$\eta_{{\rm com}}$ & $\delta_{{\rm zz}}/2\pi$ & $\eta_{{\rm zz}}$& $\Omega_{\rm L}/2 \pi$& $\Gamma$ & $\lambda_{sp}$\\
\hline
$1\,$MHz \hspace{0.1ex} & $4\,$MHz  \hspace{0.1ex} &  $127\,$kHz & \hspace{0.1ex} 0.225 & $254\,$kHz&0.229 & $811\,$kHz  & 43$\,${\rm MHz} & 280$\,$nm \\
\hline 
\hline
\end{tabular}
\label{tab_1}
\end{table}

\begin{figure}[hbt]
\includegraphics[width=0.95\textwidth]{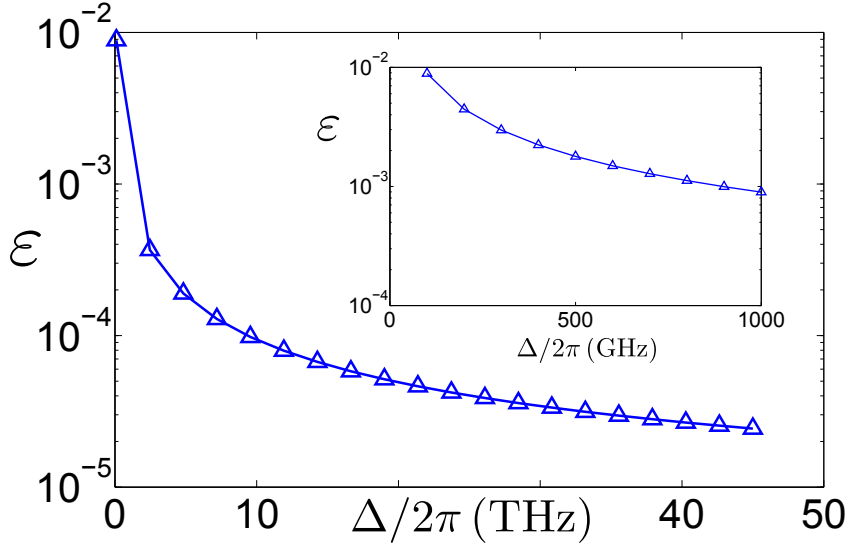}
\caption{{\bf Spontaneous emission.} Error in the generation of the Bell state $\ket{\Phi^-}$ at the expected gate time $t_{\rm g}$. Starting from the initial state $\ket{\downarrow \downarrow}$ with the phonons in the
vacuum state the time evolution includes the effects of spontaneous emission. The error is shown as a function of the Raman beam detuning. The figure shows that errors below the fault-tolerance threshold 
$\epsilon_{\rm FT} = 10^{-4}$ are only obtained for very large detunings $\Delta \sim 10\,$THz. The inset shows the error in generating $\ket{\Phi^-}$ for detunings $\Delta$ between 100$\,$GHz and 1$\,$THz. Note that for each detuning the single photon Rabi frequencies fulfill 
$|\Omega_{1,\downarrow}|=|\Omega_{2,\uparrow}|=\sqrt{|2\Delta\Omega_{\rm L}|}$ in order to maintain the gate speed.}
\label{fig_spontaneous_emission}
\end{figure}

The dynamics is then given by
\beq
\dot{\rho} = -{\rm i}[H_{\rm dss}',\rho] + \sum_{i,k} \left( L_{i,k}^{\rm eff} \rho (L_{i,k}^{\rm eff})^{\dagger} -\frac{1}{2}\{(L_{i,k}^{\rm eff})^{\dagger} L_{i,k}^{\rm eff} , \rho  \} \right)
\label{eff_master_eq_final}
\eeq
where $i=1,2$ denotes the different ions and $k=\downarrow\downarrow,\downarrow\uparrow,\uparrow\downarrow,\uparrow\uparrow$ the various scattering events. In order to investigate the effects of 
spontaneous emission exclusively, we assume both of the phonon modes to be in the ground state $\rho_{\rm vac}$, and a microwave driving of $\Omega_{\rm d} = 40\,$MHz. The phonon Hilbert spaces were
truncated at a maximum phonon number of $n_{\rm max}=11$.
For these parameters, together
with the values from Tab.~\ref{tab_1}, we integrated the master equation~\eqref{eff_master_eq_final}, starting from the state $\rho_0 = {\ketbra{\downarrow \downarrow}}\otimes \rho_{\rm vac} \otimes \rho_{\rm vac} $
until the expected gate time $t_{\rm g}\approx 63\,\mu$s. Note that we included a refocusing spin-echo pulse $\sigma_1^z \sigma_2^z$ at half the expected gate time in the dynamics in order
to avoid errors due to the fast oscillations of the applied microwave drive~\cite{driven_bermudez_gates}. 
 We expect to produce the state $\rho_{\rm t_{\rm g}} = \ketbra{\Phi^-}$ for the internal states of the ions where 
$\ket{\Phi^-}=(\ket{\downarrow \downarrow} -\ii\ket{\uparrow \uparrow})/\sqrt{2}$ is a maximally-entangled Bell state. We studied the effects of spontaneous emission by simulating the time evolution of
the gate for various detunings $\Delta$ of the Raman beams, and then computed the fidelity of producing the target state according to 
\beq
\mathcal{F}_{\ket{\Phi^-}} = {\rm tr}(\ketbra{\Phi^-}\otimes \mathds{1}_{\rm Phonons} \rho (t_{\rm g})).
\label{fidelity_computation}
\eeq
Here $\rho (t_{\rm g})$ is the state of the full system propagated until the gate time according to~\eq{eff_master_eq_final} starting from $\rho_0$. In order to illustrate the results more clearly we plot
the error $\epsilon=1-\mathcal{F}$ as function of the Raman detuning $\Delta$ in Fig.~\ref{fig_spontaneous_emission}.

We see from the figure that very large detunings of
the order of 10$\,$THz are needed to suppress spontaneous emission sufficiently to obtain errors below the fault-tolerance threshold. We note that these detunings are very large compared to the detunings that
have been used so far in two-qubit entangling gates between hyperfine qubits with cw lasers ($\Delta/2\pi\sim 100$-$250\,$GHz~\cite{didi_gate, nist_ss_gate}).  From a technological point of view, this is due to limited laser power, such that the detuning cannot be increased further without compromising the effective Rabi frequency~\eqref{eff_rabi_freq}, and thus the gate speed. From a more fundamental point of view, the conditions to nullify the ac-Stark shifts in~\eq{raman_eff_ham_ip}, impose that the detuning cannot be larger than the fine-structure splitting (e.g. $\Delta/2\pi\leq\omega_{\rm fs}/2\pi\approx 200$ GHz for $^9$Be$^+$)~\cite{wineland_errors}. However, we note that in pulsed
experiments with intense laser sources and heavier ions (e.g.$^{171}$Yb$^+$~\cite{monroe_pulsed}), the desired detunings of the order of 10$\,$THz have been implemented, while simultaneously minimizing the ac-Stark shift. An advantage of our setup is that it does not require for cancellation of the ac-Stark shift, since the associated dephasing  would be directly  suppressed by the applied strong microwave driving as was shown in~\cite{driven_bermudez_gates}. Hence, we are not bound to $\Delta\leq\omega_{\rm fs}$, and can consider higher detunings even for lighter ion species. It then seems that the only limitation to minimize the residual scattering is due to the available laser power. If this limitation cannot be overcome, it seems more promising to implement the gate in an all-microwave setup to avoid photon scattering completely.

Let us finally note that, according to the results of~\cite{raman_scattering_suppression}, Raman scattering is largely suppressed for detunings larger than the fine-structure splitting
of the excited state manifold which is about 2.75$\,$THz for $^{25}$Mg$^+$. In this case, Rayleigh scattering is the dominant spontaneous emission process. Therefore, our 
results which always include Raman and Rayleigh scattering on the same footing  represent a worst-case scenario. We should then expect that the detrimental effects of spontaneous emission will be somewhat smaller in this regime.

\subsection{Laser intensity noise}

As we have seen in the last paragraph it is desirable to operate the lasers very far detuned from the transitions they couple to. The typical detunings used for stimulated Raman transitions $\sim 100-250\,$ GHz
demand that the lasers be operated at relatively high powers in order to obtain reasonable effective laser Rabi frequencies. Unfortunately, at these high powers there will be fluctuations in the laser intensity.
These fluctuations will act as another source of noise for the gate since they lead to fluctuations of the pulse area of the applied laser pulses. In this section, we want to investigate the impact of laser 
intensity fluctuations on the gate performance quantitatively.

A fluctuating laser intensity leads to fluctuations in the laser Rabi frequency~\eqref{eff_rabi_freq}. This, in turn, leads to fluctuations in the sideband 
coupling strengths $\mathcal{F}_{in}$ as we can see from the definition~\eqref{def_sideband_coupling_strengths}. Therefore, $H_{\rm dss}'$ becomes
 \begin{equation}
 H_{\rm dss,n}' = \sum_n\delta_na_n^{\dagger}a_n+\sum_i\frac{\Omega_{\rm d}}{2}\sigma_i^x+\sum_{i,n}(\mathcal{F}_{in}(t) \sigma_i^+a_n+\text{H.c.})
\label{noisy_ham}
 \end{equation}
where $\mathcal{F}_{in}(t) = \mathcal{F}_{in} + \Delta \mathcal{F}_{in} (t) $. We assume that the sideband couplings fluctuate around some mean value $\mathcal{F}_{in}$, where the fluctuations are
described by the quantity $\Delta \mathcal{F}_{in} (t) = \ii M_{in} \Delta\Omega_{\rm L}(t) \eta_n/2$. We model the fluctuations $\Delta\Omega_{\rm L} (t)$ by a
so-called Ornstein-Uhlenbeck (O-U) process which is characterized by a diffusion constant $c$ and a correlation time $\tau$~\cite{gillespie_ou}. The O-U process is a Gaussian process and is therefore characterized by
its first and second moments
\begin{equation}
\overline{\Delta \Omega_{\rm L} (t)} = 0, \hspace{3ex}
\text{Var}\{\Delta \Omega_{\rm L} (t) \} = \frac{c \tau}{2} \left( 1- \ee^{-2t/\tau} \right).
\label{def_ou}
\end{equation}
Here the overline denotes the stochastic average. The correlation time also sets the time scale over which the noise is correlated~\cite{gillespie_ou}. We chose the correlation time $\tau=5\,\mu$s and set $ c \tau/2 = \zeta_{\rm Int}^2 \Omega_{\rm L}^2$
where $\zeta_{\rm Int} \in[5,100]\cdot 10^{-4}.$
Thus, $c$ is determined and together with Eqs.~\eqref{def_ou} the O-U process is fully characterized. Remarkably, there is an exact update formula for the O-U process
\begin{equation}
\Delta \Omega_{\rm L} (t + \Delta t) = \Delta \Omega_{\rm L} (t) \ee^{-\Delta t/\tau} + \left[ \frac{c \tau}{2} (1-\ee^{- 2 \Delta t/\tau}) \right]^{1/2} n,
\label{ou_update}
\end{equation}
where $n$ is a Gaussian unit random variable. 

\begin{figure}[hbt]
\centering
\includegraphics[width=0.95\textwidth]{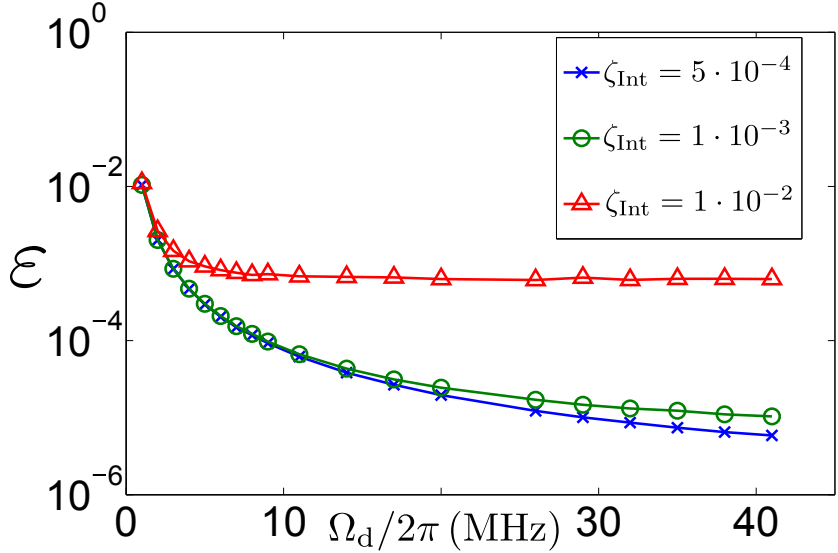}
\caption{{\bf Laser intensity noise.} The figure shows the error $\epsilon=1-\mathcal{F}$ of producing the Bell state $\ket{\Phi^-}=1/\sqrt{2}(\ket{\downarrow\downarrow} -\ii \ket{\uparrow\uparrow})$ from 
the initial state $\ket{\Psi_0} = \ket{\downarrow\downarrow}$ at the expected gate time $t_{\rm g}$. We set a truncation of $n_{\rm max} = 7$ phonons to the vibrational Hilbert spaces.
For laser intensity fluctuations of the order of $10^{-3} - 10^{-4}$ gate errors well-below the fault-tolerance threshold $\epsilon_{\rm FT} = 10^{-4}$ can be achieved.}
\label{fig_laser_intensity_noise}
\end{figure}

With our choice of parameters, we obtain an expected gate time $t_{\rm g} \approx 63\mu$s.
Starting from the initial state $\ket{\Psi_0}=\ket{\downarrow \downarrow}$, we expect to produce the Bell state $\ket{\Phi^-}=1/\sqrt{2}(\ket{\downarrow \downarrow} -\ii \ket{\uparrow \uparrow})$ at the gate time. With the update 
formula at hand we numerically integrated the noisy Hamiltonian~\eq{noisy_ham} incorporating the laser intensity fluctuations until the expected gate time, and computed the fidelity of producing $\ket{\Phi^-}$ according to
\beq
\mathcal{F}_{\ket{\Phi^-}} = {\rm tr}(\ketbra{\Phi^-}\otimes \mathds{1}_{\rm Phonons} \ketbra{\Psi (t_{\rm g})} )
\eeq
where $\ket{\Psi (t_{\rm g})}$ denotes the state of the full system propagated in time by the noisy Hamiltonian~\eqref{noisy_ham}. Once again we introduced a refocusing
spin-echo pulse in the $\sigma^z$ basis at half the expected gate time.
 In order to solely capture the errors introduced by the laser intensity fluctuations, we assumed the phonons
to be in the ground state initially. In Fig.~\ref{fig_laser_intensity_noise} the results of our simulations are illustrated. We plot the error $\epsilon = 1- \mathcal{F}$ of producing the Bell state $\ket{\Phi^-}$ at
the expected gate time. As it can be clearly seen from the figure, errors well-below the fault-tolerance threshold $\epsilon_{\rm FT}=10^{-4}$ can be obtained for large drivings and relative intensity fluctuations on the
order of $10^{-3}$ or smaller. However, for relative fluctuations of $10^{-2}$ the gate cannot attain errors below the FT.

\section{Conclusions and outlook}\label{chapter_conclusions}

In this manuscript, we have investigated the effects of spontaneous emission and laser intensity fluctuations on the performance of driven geometric phase gates as introduced in~\cite{driven_bermudez_gates}.
We have found that laser intensity fluctuations of the order of $\sim 10^{-3}$ allow for errors below the stringent fault-tolerance threshold of $10^{-4}$. Spontaneous emission turns out to be 
the more substantial source of infidelity. Only for detunings of the order of $\sim 10\,$THz spontaneous emission is suppressed sufficiently to obtain errors below $10^{-4}$. In combination with the results in~\cite{driven_bermudez_gates}, this shows that the driven geometric phase gate is robust (i.e. it can beat the FT) for several sources of noise, namely thermal ion motion, dephasing noise, laser phase drifts, laser intensity fluctuations, and residual photon scattering.
Moreover, strong-driving entangling gates can be implemented with an always-on sympathetic cooling to overcome the errors due to motional heating of the ions~\cite{last_ref}.

\appendix


\begin{thebibliography}{10}
\newcommand{\enquote}[1]{``#1''}
\expandafter\ifx\csname url\endcsname\relax
  \def\url#1{{#1}}\fi
\expandafter\ifx\csname urlprefix\endcsname\relax\def\urlprefix{}\fi

\bibitem{driven_bermudez_gates}
A.~Lemmer, A.~Bermudez, and M.~B. Plenio, New J. Phys. {\bf 15}, 083\,001
  (2013).

\bibitem{NC2010}
M.~A. Nielsen and I.~L. Chuang, {\em Quantum Information and Quantum
  Computation\/} (Cambridge University Press, 2010).

\bibitem{two_qubit_gates_universal}
D.~DiVincenzo, Phys. Rev. A {\bf 51}, 1015 (1995).

\bibitem{qc_review}
{\it See} T. Ladd, F. Jelezko, R. Laflamme, Y. Nakamura, C. Monroe, and J. L. OBrien, Nature {\bf 464,} 45 (2010), {\it and references therein}.

\bibitem{knill}
E.~Knill, Nature {\bf 463}, 441 (2010).

\bibitem{qc_ions}
{\it See} H. H\"{a}ffner, C. F. Roos, and R. Blatt, Phys. Rep. {\bf 469,} 155
(2008), {\it and references therein.}

\bibitem{1_qubit_mw_gates}
K. R. Brown, A. C. Wilson, Y. Colombe, C. Ospelkaus, A. M.
Meier, E. Knill, D. Leibfried, and D. J. Wineland, Phys. Rev. A
{\bf 84,} 030303(R) (2011).

\bibitem{zoller_cirac}
J.~I. Cirac and P.~Zoller, Phys. Rev. Lett. {\bf 74}, 4091 (1995); F. Schmidt-Kaler, H. H\"{a}ffner, M. Riebe, S. Gulde, G. P. T. Lancaster, T. Deuschle, C. Becher, C. F. Roos, J. Eschner, and R.
Blatt, Nature {\bf 422,} 408 (2003).

\bibitem{ms_gate1}
A.~S{\o}rensen and K.~M{\o}lmer, Phys. Rev. Lett. {\bf 82}, 1971 (1999); C. A. Sackett, D. Kielpinski, B. E. King, C. Langer, V. Meyer, C. J. Myatt, M. Rowe, Q. A. Turchette, W. M. Itano, D. J.
Wineland, and C. Monroe, Nature {\bf 404,} 256 (2000); J. Benhelm, G. Kirchmair, and R. Blatt, Nat. Phys. {\bf 4,} 463
(2008).

\bibitem{didi_gate}
G. J. Milburn, S. Schneider, and D.F.V. James, Fortschr. Phys.
{\bf 48,} 801(2000); D.~Leibfried, B.~DeMarco, V.~Meyer, D.~Lucas, M.~Barrett, J.~Britton, W.~M.
  Itano, B.~Jelenkovic, C.~Langer, T.~Rosenband, and D.~J. Wineland, Nature
  {\bf 422}, 412 (2003).


\bibitem{far_detuned_driven_geometric_phase_gate}
A. Bermudez, P. O. Schmidt, M. B. Plenio, and A. Retzker,
Phys. Rev. A {\bf 85,} 040302(R) (2012).

\bibitem{nist_ss_gate}
T.~R. Tan, J.~P. Gaebler, R.~Bowler, Y.~Lin, J.~D. Jost, D.~Leibfried, and
  D.~J. Wineland, Phys. Rev. Lett.  {\bf 110,} 263002 (2013).

\bibitem{static_gradients}
F.~Mintert and C.~Wunderlich, Phys. Rev. Lett. {\bf 87}, 257\,904 (2001); A. Khromova, Ch. Piltz, B. Scharfenberger, T. F. Gloger, M. Johanning, A. F. Varon, and Ch. Wunderlich, Phys. Rev. Lett. {\bf 108,} 220502 (2012).
  
\bibitem{oscillating_gradients}
C.~Ospelkaus, C.~E. Langer, J.~M. Amini, K.~R. Brown, D.~Leibfried, and D.~J.
  Wineland, Phys. Rev. Lett. {\bf 101}, 090\,502 (2008); C. Ospelkaus, U. Warring, Y. Colombe, K. R. Brown, J. M. Amini, D. Leibfried, and D. J. Wineland, Nature {\bf 476,} 181 (2011).
	
\bibitem{plenio_lower_bounds}
M.~B. Plenio and P.~L. Knight, Phys. Rev. A {\bf 53}, 2986 (1996).

\bibitem{plenio_lower_bounds_proc}
M.~B. Plenio and P.~L. Knight, Proc. R. Soc. Lond. A {\bf 453}, 2017 (1997).

\bibitem{sorensen_reiter}
F.~Reiter and A.~S{\o}rensen, Phys. Rev. A {\bf 85}, 032\,111 (2012).


	\bibitem{master_eq_review_susana}
	A. Rivas and S.~F. Huelga, {\em Open Quantum Systems - An Introduction\/}, SpringerBriefs in Physics (2012)
  
  \bibitem{comment_ac_ss}
  In experimental situations, one must sum over all possible excited levels, such that the cancellation of  the ac-Stark shift is more involved, yet possible~\cite{wineland_errors}.
  
  \bibitem{wineland_errors}
  D. J. Wineland, M. Barrett, J. Britton, J. Chiaverini, B. DeMarco, W. M. Itano, B. Jelenkovic, C. Langer, D. Leibfried, V. Meyer, T. Rosenband and T. Schaetz, Phil. Trans. R. Soc. Lond. A {\bf 361,}
1349 (2003).

\bibitem{james}
D.~F.~V. James, Appl. Phys. B {\bf 66}, 181 (1998).

\bibitem{light_shift_gates}
D. Jonathan, M.~B. Plenio and P.~L. Knight, Phys. Rev. A {\bf 62}, 042307 (2000).

\bibitem{light_shift_gates_thermal}
D. Jonathan and M.~B. Plenio, Phys. Rev. Lett. {\bf 87}, 127901 (2001).



\bibitem{strong_driving_cavities_solano}
E. Solano, G. S. Agarwal, and H. Walther, Phys. Rev. Lett. {\bf 90,} 027903 (2003).

\bibitem{strong_driving_cavities_zheng}
 S.-B. Zheng, Phys. Rev. A {\bf 66,} 060303(R) (2002).


\bibitem{monroe_pulsed}
J.~Mizrahi, C.~Senko, B.~Neyenhuis, K.~Johnson, W.~Campbell, C.~Conover, and
  C.~Monroe, Phys. Rev. Lett. {\bf 110}, 203\,001 (2013).

\bibitem{raman_scattering_suppression}
R.~Cline, J.~Miller, M.~Matthews, and D.~Heinzen, Opt. Lett. {\bf 19}, 207
  (1994).

\bibitem{gillespie_ou}
D.~T. Gillespie, Am. J. Phys. {\bf 64}, 225 (1996).

\bibitem{last_ref}
A. Bermudez, T. Schaetz, and M.B. Plenio, Phys. Rev. Lett. {\bf 110}, 110502 (2013).

\end{thebibliography}

\end{document}